\documentclass{article}
\usepackage{graphicx}
\usepackage{epsfig}
\usepackage{amsmath}
\usepackage{amssymb}
\def\bea{\begin{eqnarray}}
\def\eea{\end{eqnarray}}
\def\be{\begin{equation}}
\def\ee{\end{equation}}
\def\p{\partial}
\def\nn{\nonumber}

\begin{document}

\baselineskip=16pt
\begin{titlepage}
\setcounter{page}{0}
\begin{center}

\vspace{0.5cm}
 {\Large \bf Note on DBI dynamics of Dbrane Near NS5-branes}\\
\vspace{10mm} {\large Bin Chen\footnote{e-mail
 address:bchen@itp.ac.cn}}\\
\vspace{6mm} {\it Interdisciplinary Center of
Theoretical Studies, Chinese Academy of Science,\\
 P.O. Box 2735, Beijing 100080, P.R. China}\\
 \vspace{10mm} {\large
 Bo Sun\footnote{e-mail address: sunbo@itp.ac.cn}}\\
 \vspace{6mm} {\it
      Institute of Theoretical Physics, Chinese Academy of
      Sciences,\\
       P.O. Box 2735,
      Beijing 100080, P.R. China\\}
\vspace{6mm} {\it  Graduate School of the Chinese Academy of
 Science, Beijing 100080, P.R. China\\}

\vspace*{5mm} \normalsize
\smallskip
\medskip
\smallskip
\end{center}
\vskip0.6in \centerline{\large\bf Abstract} \vspace{1cm} {In this
note, we investigate the homogeneous radial dynamics of (Dp,
NS5)-systems without and with one compactified transverse
direction, in the framework of DBI effective action. During the
homogeneous evolution, the electric field on the D-brane is always
conserved and the radial motion could be reduced to an
one-dimension dynamical system with an effective potential. When
the Dp-brane energy is not high, the brane moves in a restricted
region, with the orbits depending on the conserved energy, angular
momentum through the form of the effective potential. When the
Dp-brane energy is high enough, it can escape to the infinity. It
turns out that the conserved angular momentum plays an interesting
role in the dynamics. Moreover, we discuss the gauge dynamics
around the tachyon vacuum and find that the dynamics is very
reminiscent of the string fluid in the rolling tachyon case.}

\vspace*{2mm}

\end{titlepage}

\section{Introduction}

The Dirac-Born-Infeld (DBI) field theory encodes the dynamics of
D-brane\cite{Dai}. It has been proved to be a very powerful tool
to study various aspects of the D-brane dynamics. In \cite{Sen1},
A. Sen proposed a DBI-like effective field theory to study the
tachyon dynamics on the non-BPS brane. Very surprisingly, the
effective field theory describe the tachyon condensation quite
well\cite{Sen2}. This fact leads to a new kind of open-closed
string duality\cite{Sen4}.

Very recently, it has been found that the radial behavior of
Dp-brane near NS5-branes is very similar to the rolling
tachyon\cite{Kutasov1}. In the near throat region, the classical
motion of D-brane is a hair-pin brane in CHS theory, which could
be studied in the framework of BCFT. In \cite{Nakayama}, the
boundary state of the supersymmetric hairpin brane has been
constructed and in \cite{Nakayama, Sahakyan} the closed string
radiation of the D-brane near NS5-branes has been discussed. In
\cite{Chen1}, taking into account of the constant electromagnetic
field, the radial dynamics and closed string radiation has been
investigated. (see also \cite{Rey1}) It was found that due to the
constant electric field, the closed string radiation without
winding is finite. Nevertheless, the emission of the closed string
with windings dominate the radiation and is divergent. These facts
are reminiscent of the closed string radiation of the rolling
tachyon in the presence of constant NS $B_{\mu\nu}$ field.
Therefore, the (Dp, NS5)-system gives another interesting
laboratory to study the string theory in a time-dependent
background. Other related discussions can be found in \cite{Rey2}.

Furthermore, in \cite{Kutasov2} D. Kutasov studied the (Dp, NS5)
system with one compactified transverse direction (which we call
(Dp, NS5)' system) and noticed that certain (Dp, NS5)
configuration has kink solution and behaves like a non-BPS brane
in type II theory. Unfortunately, the CFT analysis on this
configuration seems to be out of reach. In this note, we take the
philosophy that DBI effective action catch most of the tachyon
dynamics and investigate the DBI dynamics of (Dp, NS5)-systems. We
will concentrate on the homogeneous evolution in which case the
radial motion could be directly related to the rolling tachyon.
However, the introduction of the angular momentum into the system
make the dynamics much richer.  (see \cite{Branium} for the
similar discussions in Dp-anti-Dp system)

Another interesting issue is the gauge dynamics of the tachyon
condensation in the DBI effective action. In the rolling tachyon
case, it has been shown that
 two components of pressureless fluid survive the tachyon condensation:
 one is conserved electric flux lines, named string fluid; the
 other is the dust-like pressureless tachyon matter\cite{Pil1,Pil2}. Around the
 tachyon vacuum, the gauge dynamics could be well described by
 a set of fluid equations with integrability
 conditions\cite{Pil3}. Further investigation shows that such
 string fluid may have a closed string interpretation\cite{Sen3,Sen4,Pil4}.
 We will study this issue in the (Dp, NS5) geometric tachyon
 configurations.

We will revisit (Dp, NS5)-system with gauge field fluctuations and
angular momentum in section 2. We study the radial dynamics of
(Dp, NS5)'-system in section 3. Discussions and conclusions will
be in section 4.

\section{The Radial Dynamics of (Dp, NS5)-system}

In this section we use the effective DBI action  to analyze the
Dp-brane dynamics near the NS5-branes. The tension of a NS5-brane
$\sim 1/g_s^2$ while the tension of a $Dp$-brane $\sim 1/g_s$, so
it is natural to take the NS5-branes' supergravity solution as the
background when the string coupling is weak.

The coordinates on the world-volume of $k$ coincident NS5-branes
are $x^\mu$, $\mu=0,1,...5$, and we use $x^n,n=6,7,8,9$ to label
the four transverse dimensions. Let the Dp-brane be parallel to
NS5-branes and let the world-volume of Dp-brane lie along $x^0,
\cdots x^p$ with $2 \le p < 5$.  Such a system breaks
supersymmetry completely and is unstable.

Setting $r^2=\displaystyle{\sum_{n=6}^{9} x^nx^n}$, the low energy
supergravity solution of NS5-branes is
\begin{eqnarray}\label{metric}
ds^2&=&dx^\mu dx_\mu+H(r)dx^ndx_n\equiv g_{MN}dx^Mdx^N\nonumber\\
\frac{g_s(\Phi)}{g_s}&=&\exp(\Phi-\Phi_0)=\sqrt{H(r)}\nonumber\\
H_{mnp}&=&-\epsilon_{qmnp}\partial^q \Phi,\nonumber \\
H(r)&=&1+\frac{kl_s^2}{r^2},
\end {eqnarray}
where $H_{mnp}$ is the NS 2-form field strength, $g_s$ is the
asymptotic string coupling, and $l_s$ is the string length unit.

\subsection{Dynamics without angular momentum}

Since there is an $SO(4)$ rotational symmetry for the four
transverse dimensions, the angular momentum is conserved. Let's
temporally ignore the angular momentum and focus on the radial
motion of the bounded system first. The dynamics of Dp-brane in
the NS5-branes background is well
 described by the DBI action
\begin{equation}\label{action}
S_p=-\tau_p\int d^{p+1}x
e^{-(\Phi-\Phi_0)}\sqrt{-det(G_{\mu\nu}+B_{\mu\nu}+F_{\mu\nu})}
\end{equation}
where $\tau_p$ is the asymptotic tension of the Dp-brane
\begin{equation}\label{tau}
\tau_p\sim\frac{1}{g_sl_s^{p+1}}
\end{equation}
There is a U(1) gauge freedom which can be used to set $A_0=0$. We
are free to set $B_{\mu\nu}=0$ due to the gauge transformation and
we have:
\begin{eqnarray}
-Det(G+F)=Det(h)(1-H\dot{r}^2)-E^+_iD_{ik}E^-_k\\
E^{\pm}_i=F_{0i}\pm H\dot{r}\partial_i{r}=E_i\pm
H\dot{r}\partial_i{r}\\
h_{ij}=\delta_{ij}+F_{ij}+H\partial_ir\partial_jr\\
D_{ij}=(-1)^{i+j}\Delta_{ji}(h)=Det(h)h^{-1}_{ij}
\end{eqnarray}
Then the action is written as
\begin{eqnarray}
-\tau_p\int dt
\frac{1}{\sqrt{H}}\sqrt{Det(h)(1-H\dot{r}^2)-E^+_iD_{ik}E^-_k}
\end{eqnarray}
From this we can evaluate the canonical momentum conjugate to
$A_{i}$ and $r$:
\begin{eqnarray}
\mathbf{\Pi}_r=\frac{\tau_p}{\sqrt{H}\sqrt{Det(h)(1-H\dot{r}^2)-E^+_iD_{ik}E^-_k}}(Det(h)H\dot{r}-\frac{E^+_kD_{ki}-D_{ik}E^-_k}{2}H\partial_ir)\\
\mathbf{\Pi}_e^i=\frac{\tau_p}{\sqrt{H}\sqrt{Det(h)(1-H\dot{r}^2)-E^+_iD_{ik}E^-_k}}(\frac{E^+_kD_{ki}+D_{ik}E^-_k}{2})
\end{eqnarray}
And the conserved Hamiltonian is
\begin{eqnarray}\label{hamiltonian}
\mathcal{H}&=&\frac{\tau_pDet(h)}{{\sqrt{H}\sqrt{Det(h)(1-H\dot{r}^2)-E^+_iD_{ik}E^-_k}}}\nonumber\\
&=&\sqrt{|\vec{\mathbf{\Pi}}_e|^2+\frac{1}{H}\mathbf{\Pi}_r^2+H(\mathbf{\Pi}_e^i\partial_ir)^2+|\vec{\mathcal{P}}|^2+\frac{Det(h)}{H}}\\
\mathcal{P}_i&=&-F_{ik}\mathbf{\Pi}_e^k-\partial_ir\mathbf{\Pi}_r
\end{eqnarray}
In the above $\mathcal{P}$ is the conserved Noether charge
associated with the spatial translation along the world volume.

As remarked in \cite{Kutasov1}, after the field redefinition, one
may treat $r$ as the rolling tachyon field and $1/\sqrt{H}$ as the
tachyon potential. As the geometric tachyon field $r\rightarrow
0$, the tachyon potential tends to zero, namely as D-brane falls
down close to the NS5-branes, $H\rightarrow\infty$. From
(\ref{hamiltonian}),   the last term in it vanishes, and the term
$H(\mathbf{\Pi}_e^i\partial_ir)^2$ dominates. So the energy
conservation requires that $\partial_i r\approx 0$ so the Dp-brane
inclines to homogenously evolve in the near throat region.




Now we treat the motion of the Dp-brane as homogeneous radial
evolution, that is all the world volume fields are only functions
of $X^0$ or $t$, so the only non-vanishing components of
$F_{\mu\nu}$ is $F_{0i}=-F_{i0}=\dot{A}_i=e_i$. The pull-back
metric and the world-volume gauge field is
\begin{equation}
\mathbf{G_{\mu\nu}}= \left(\begin{array}{cccc}
-1+H\dot{r}^2&0&0&0\\
0&1&0&0\\
0&0&1&0\\
0&0&0&1\\
\end{array}\right)\hspace{5ex}
\mathbf{F_{\mu\nu}}= \left(\begin{array}{cccc}
0&e_1&e_2&e_3\\
-e_1&0&0&0\\
-e_2&0&0&0\\
-e_3&0&0&0\\
\end{array}\right).
\end{equation}
And the DBI action reads
\begin{eqnarray}
S_p=-\tau_p\int d^{p+1}x \frac{1}{\sqrt{H}}\sqrt{1-H\dot{r}^2-e^2}
=-\tau_pV\int dt \frac{1}{\sqrt{H}}K\nonumber\\
K=\sqrt{1-H\dot{r}^2-e^2}\,\,\,, e^2=\sum_{i=1}^{p}e_i^2
\end{eqnarray}
where $V$ is the volume of the D-brane.

The canonical momentum density conjugate to $r$ and $A_i$ are as
follows:
\begin{eqnarray}
\mathbf{\Pi}_r=\tau_p\frac{\sqrt{H}}{K}\dot{r}\nonumber\\
\mathbf{\Pi}_e^i=\tau_p\frac{e_i}{\sqrt{H}K}=n_i
\end{eqnarray}
and the conserved Hamiltonian is
\begin{eqnarray}
\mathcal{H}=\mathbf{\Pi}_r\dot{r}+\mathbf{\Pi}_e^ie_i-\mathcal{L}_{DBI}=\frac{\tau_p}{\sqrt{H}K}=E
\end{eqnarray}

  We can obtain  the strength tensor $T_{\mu\nu}$ and NS source tensor
$S_{\mu\nu}$ as in \cite{Chen1}
\begin{eqnarray}
\delta S&=&-\frac{\tau_p}{2}e^{-(\Phi-\Phi_0)}\sqrt{-det(\mathbf{G}+\mathbf{B})}(\mathbf{G}+\mathbf{B})^{\mu\nu}(\delta g_{\mu\nu}+\delta b_{\mu\nu})\nonumber\\
&=&-\frac{\tau_p}{2}\frac{1}{\sqrt{H(r)}}\sqrt{-det(\mathbf{G}+\mathbf{B})}(\mathbf{G}+\mathbf{B})^{\mu\nu}(\delta
g_{\mu\nu}+\delta b_{\mu\nu})\nonumber
\end{eqnarray}
The result is:
\begin{eqnarray}
\mathbf{T_{\mu\nu}}=\frac{\tau_p}{\sqrt{H}K}\left(\begin{array}{cccc}
1&0&0&0\\
0&e_1^2-K^2&-e_1e_2&-e_1e_3\\
0&-e_1e_2&e_2^2-K^2&-e_2e_3\\
0&-e_1e_3&-e_2e_3&e_3^2-K^2\\
\end{array}\right)\,,
\mathbf{S_{\mu\nu}}=\frac{\tau_p}{\sqrt{H}K}\left(\begin{array}{cccc}
0&-e_1&-e_2&-e_3\\
e_1&0&0&0\\
e_2&0&0&0\\
e_3&0&0&0\\
\end{array}\right).
\end{eqnarray}

The equation of motion can be written as  an one-dimensional
problem effectively:
\begin{eqnarray}
\dot{r}^2=-V_{eff}=\frac{1}{H}-\frac{\tau_p^2+Hn^2}{H^2E^2}
\end{eqnarray}
We can classify the motion of the D-brane according to the
behavior of $V_{eff}$. Set $a=\frac{\tau_p^2}{E^2}$,
$b=\frac{n^2}{E^2}\leq 1 $, then
\begin{eqnarray}
V_{eff}=\frac{a}{H^2}+\frac{b-1}{H}.
\end{eqnarray}
We have
\def\labelenumi{\theenumi}
\def\theenumi{\roman{enumi}.}
\begin{enumerate}


\item when $\frac{1-b}{a}<1$, or $E^2< n^2+\tau_p^2$, the D-brane
move between $r=0$ and $r=r_{max}$ where
 \be
 r_{max}=\sqrt{k}\sqrt{\frac{E^2-n^2}{n^2+\tau_p^2-E^2}}l_s
 \ee

\item when $\frac{1-b}{a}\geq 1$, or $E^2\geq n^2+\tau_p^2$, the
brane could escape to infinity.
\end{enumerate}

In the near throat region, one has $H=\frac{kl_s^2}{r^2}$ and one
can solve the equations of motions and get a solution as the
modified hair-pin brane.

In the above we set the conserved quantity $\mathbf{\Pi}_e^i$ to
$n_i$, which is the electric flux long the direction $x^i$ on the
D-brane. In the usual discussion of the DBI action with gauge
field, the conserved quantity is only the electric flux $n_i$
rather than the electric field $e_i$\cite{Pil1}. However as a
direct consequence of homogeneity, the electric field $e_i$ is
also conserved in our case:
\begin{eqnarray}
e_i=\frac{n_i}{E}
\end{eqnarray}
so that the problem is reduced to the case studied in some details
in \cite{Chen1}. The energy conservation requires that when the
D-brane approaches very closely to the NS5-branes, the geometric
tachyon reaches its vacuum and $H\rightarrow \infty$, so one has
$K=0$, i.e. $H\dot{r}^2=1-e^2$. When the electric field takes the
critical value $e=1$, the open-string degrees of freedom on the
D-brane decouples from the bulk closed string modes, the geometric
tachyon does not roll and there is no closed string radiation from
the D-brane and no tachyon matter. When $e<1$, the behavior of the
D-brane in the near throat region of NS5-background turns out to
be a modified hairpin solution. The presence of the electric field
slows down the rolling of the geometric tachyon and in the end of
the tachyon condensation, besides the usual tachyon matter, there
exist the fluid of electric flux lines. This could be seen from
the partition of the energy at the vacuum:
 \be
 \mathcal{H}=E=\sqrt{\mathbf{\Pi}_e^2+\frac{\mathbf{\Pi}_r^2}{H}}
 \ee
 where
 \be
 \mathbf{\Pi}_e^2=\sum_in_i^2=e^2E^2, \hspace{5ex}
 \frac{\mathbf{\Pi}_r^2}{H}=(1-e^2)E^2
 \ee
are the flux energy and kinetic energy respectively. The partition
of the energy is reminiscent of the vacua $\dot{T}^2+e^2=1$ in the
rolling tachyon case\cite{Pil2}. Therefore, when Dp-brane fall
towards the NS5-branes, it loses energy by  closed string
radiation but  the constant electric field slows its motion and
survive the tachyon condensation as the string fluid.

One may investigate the fluctuations around the tachyon vacuum
from the Hamiltonian dynamics\cite{Pil2}. Under the identification
 \be
 \p_iT=\sqrt{H}\p_ir, \hspace{5ex} \mathbf{\Pi}^T=
 \frac{\mathbf{\Pi}_r}{\sqrt{H}},
 \ee
 and take $T$ as a component of gauge field along an imagined
 direction
 \be
 A_T=T,
 \ee
we may rewrite the Hamiltonian (\ref{hamiltonian}) at the end of
geometric tachyon condensation as
 \be\label{end}
 \mathcal{H}=\sqrt{\mathbf{\Pi}^I\mathbf{\Pi}^I+(F_{IJ}\mathbf{\Pi}^J)^2},
 \ee
 where $\p_T \equiv 0$ on any object and $I, J=1,2,\cdots p$ and
 $T$. This relation is exactly the same as the one in the rolling
 tachyon case. Correspondingly, one has the same Hamiltonian
 equations of motions. And the discussion of the fluctuations around the constant
 background $\mathbf{\Pi}^M_0$ follows the same line. It turns out
 the dynamical fluctuations obey the equation
 \be
 (\p_t^2-\sum_{i=1}^qe^2_i\p_i^2)\delta\mathbf{\Pi}^M=0,
 \ee
where we assume that the electric fields appear in $q$ directions.
After proper rescaling $x_i^\prime=\frac{x_i}{e_i}$, the above
equation could be transformed to the wave function in $q$ spatial
dimensions. The general solution to the equation is
 \be
 f_L(t-\vec{k}\cdot \vec{r}^\prime,x_j)+f_R(t+\vec{k}\cdot \vec{r}^\prime,x_j)
 \ee
 where $\vec{k}$ is any unit vector and $x_j$ is the directions without electric field. When all $e_i$ vanish, the
 solution shows the Carrollian behavior discussed in \cite{Pil2}.
 When $q=1$, the solution reduces to the one where $p$ degrees of
 freedom propagate along $\vec{e}$ at speed $e$.

From the above discussion, it is obvious that after field
definition, the geometric tachyon condensation problem in (Dp,
NS5)-system could be identified with the  rolling tachyon problem
on a non-BPS brane. Such identification works not only in the pure
tachyon case, it also works in the case when the tachyon coupled
to the worldvolume gauge fields. In \cite{Pil3}, the classical
dynamics of the tachyon field coupled to the gauge field has been
studied carefully, in the framework of Hamiltonian dynamics. It
could be shown that the canonical equations of motions and the
conservation of the energy-momentum tensor give us the fluid
equations of motions
 \bea
 \p_0 n^I+v^i\p_i n^I&=&n^i\p_i v^I \nn\\
 \p_0v^I+v^i\p_iv^I&=&n^i\p_in^I \label{fluid}
 \eea
 with the integrability condition
 \be \label{inte}
 v^i=-F_{ij}n^j-n^T\p_iT, \hspace{5ex}v^T=\p_iTn^i,
  \ee
where
 \be
 n^i=\frac{\mathbf{\Pi}^i}{\mathcal{H}},\hspace{5ex}
 n^T=\frac{\mathbf{\Pi}^T}{\mathcal{H}}\ee
 \be
 v^i=\frac{F_{ij}\mathbf{\Pi}^j}{\mathcal{H}},\hspace{5ex}
 v^T=\frac{\p_iT\mathbf{\Pi}^i}{\mathcal{H}}
 \ee
 satisfying
 \be
 n^In^I+v^Iv^I=1, \hspace{5ex}n^Iv^I=0.
 \ee
The various classical solutions representing the distribution of
the tachyon matter and string fluid can be obtained as in
\cite{Pil3}. Even though there is tight coupling between string
fluid and tachyon matter fluid, the distribution of the two fluid
components in the static configuration, which has $\p_0=0,
F_{ij}\mathbf{\Pi}^j=0$, is almost independent with each other and
the only requirement is that the static distribution of the
tachyon matter must be stay along the string fluid direction. For
example, the case we discussed above corresponds to the electric
flux densities along $n^i$ which have electric fields, and
nonvanishing tachyon matter $\mathbf{\Pi}$. Moreover, if there
exist a momentum density, the evolution of the geometric tachyon
is not homogeneous. For instance, one may allow a boost along
$x^1$ to induce a momentum $v^1=-n^T\p_1T$ without turning on
magnetic field. This implies that $\p_1T\neq 0$. From the field
redefinition, approximately one has
 \be
 r=\exp\left(-\frac{v^1}{\sqrt{(1-e^2)k}l_s}x^1\right).
 \ee
  In other words, the inhomogeneous
evolution of the radial direction is suppressed in the near
horizon region.

\subsection{Dynamics with angular momentum}

Now we consider the motion of the bounded system with angular
momentum. By using the transverse $SO(4)$ symmetry, the orbit can
be confined on the $X^6,X^7$ plane. Set
\begin{eqnarray}
X^6=R\cos\theta\,\,, \hspace{3ex}X^7=R\sin\theta,
\end{eqnarray}
the DBI action then reads
\begin{eqnarray}
S_p=-\tau_p\int d^{p+1}x
\frac{1}{\sqrt{H}}\sqrt{1-H(\dot{R}^2+R^2\dot{\theta}^2)-e^2}=-\tau_pV\int
dt \frac{K}{\sqrt{H}}
\end{eqnarray}
where $K=\sqrt{1-H(\dot{R}^2+R^2\dot{\theta}^2)-e^2}$.

Now the canonical momentum conjugate to $R$, $\theta$, $A_i$ are
\begin{eqnarray}
\mathbf{\Pi}_R=\frac{\delta S_p}{\delta \dot{R}}=\tau_p\frac{\sqrt{H}\dot{R}}{K}\nonumber\\
\mathbf{\Pi}_\theta=\frac{\delta S_p}{\delta \dot{\theta}}=\tau_p\frac{\sqrt{H}}{K}R^2\dot{\theta}=L\nonumber\\
\mathbf{\Pi}_e^i=\frac{\delta S_p}{\delta
\dot{A_i}}=\tau_P\frac{e_i}{\sqrt{H}K}=n_i
\end{eqnarray}
and the Hamiltonian is
\begin{eqnarray}
\mathcal{H}=\mathbf{\Pi}_R\dot{R}+\mathbf{\Pi}_\theta\dot{\theta}+\mathbf{\Pi}_e^ie_i-\mathcal{L}_{DBI}=E
=\frac{\tau_p}{\sqrt{H}K}
\end{eqnarray}
In the above we have set the conserved quantity
$\mathbf{\Pi}_\theta$, $\mathbf{\Pi}_e^i$, $\mathcal{H}$ as $L$,
$n_i$, and $E$ respectively.

Similarly,  the equation of motion of $R$ is reduced to an
one-dimensional problem:
\begin{eqnarray}
\dot{R}^2=-V_{eff}=\frac{1}{H}-\frac{1}{H^2E^2}(\tau_p^2+\frac{L^2}{R^2}+n^2H)
\end{eqnarray}
where $n^2=\displaystyle\sum_{i=1}^{p}n_i^2$.

When $R$ is large compared to $l_s$, one could approximate
$H^{-1}$ and get
 \be
 -V_{eff}=(1-e^2-\frac{\tau^2_p}{E^2})+
 \frac{1}{R^2}((\frac{2\tau_p^2}{E^2}-1+e^2)kl^2_s-\frac{L^2}{E^2}).
 \ee
Therefore, it is easy to figure out that
\def\labelenumi{\arabic{enumi}.}
\begin{enumerate}
\item if $E^2< n^2+\tau^2_p$, and
 \be
 \frac{L^2}{E^2}< (\frac{2\tau^2_p}{E^2}-1+e^2)kl^2_s,
 \ee
 then $R\in [0, 1/u_0]$, where $u_0$ is the positive root of
 $V_{eff}=0$; if $L$ is larger, the radial motion is frozen.

 \item if $n^2+2\tau^2_p>E^2\geq n^2+\tau^2_p$, D-brane move between
 $R=0$ and infinity;
\item if $E^2>n^2+2\tau^2_p$ or $n^2+2\tau^2_p>E^2\geq
n^2+\tau^2_p$ but $L$ is large, there is a potential barrier for
the
 D-brane to reach $R=0$, $R \in [1/u_0, \infty]$, where $u_0$ is the positive root of
 $V_{eff}=0$.
 \end{enumerate}
 Obviously, in order to have bounded solution, we need to require $E^2<
 n^2+\tau^2_p$.

When the D-brane reaches the near horizon region, one has
 \be
 \dot{R}^2=(\frac{1-e^2}{kl^2_s}-\frac{L^2}{k^2l^4_sE^2})R^2-\frac{\tau^2_p}{k^2l^4_sE^2}R^4
 \ee
and the solution is
\begin{eqnarray}\label{cosh}
r=\frac{\beta}{\cosh\alpha t}
\end{eqnarray}
where
 \be
 \alpha^2=\frac{1}{kl_s^2}(1-e^2-\frac{L^2}{kl_s^2E^2})\,\,,\,\,\,\,\beta^2=\frac{kl_s^2E^2}{\tau_p^2}(1-e^2-\frac{L^2}{kl_s^2E^2}).
\ee In order to have bounded solution, beside requiring
$E^2<n^2+\tau_p^2$, one has also require
 \be\label{ang}
 \frac{L}{E}< \sqrt{(1-e^2)k}l_s.
 \ee

The existence of the angular momentum will slow down the falling
of the Dp-brane to the NS5-branes, just like the role played by
the constant electric field.

To study the remnants after tachyon condensation, we rewrite the
Hamiltonian as
 \be
 \mathcal{H}=\sqrt{\mathbf{\Pi}^2_e+\frac{\mathbf{\Pi}^2_R}{H}+\frac{\mathbf{\Pi}_\theta^2}{R^2H}+\frac{\tau_p^2}{H}}.
 \ee
 At the end of tachyon condensation, $H\rightarrow \infty$, the
 potential vanishes.  The novel feature here is that the angular
 momentum do contribute to the energy since $R^2H=kl_s^2$. Besides the
 string fluid and tachyon matter, the angular momentum survives the tachyon condensation. The
 kinetic energy of geometric tachyon is
 \be
 \frac{\mathbf{\pi}^2_R}{H}=(1-e^2)E^2-\frac{L^2}{kl^2_s},
 \ee
 which implies (\ref{ang}). In this case, due to the existence of
 the angular momentum, the electric field take its critical value
 at
 \be
 e_c=1-\frac{L^2}{kl^2_sE^2}
 \ee
 when the radial motion is frozen.

It is remarkable that even with the existence of the angular
momentum, the remnants of the tachyon condensation is still
 pressureless fluid. If the D-brane initially has nonzero angular
 momentum, it will spirally fall to the NS5-branes. During the
 falling, it radiate closed string modes and lose energy. The
 electric field is still constant during homogeneous evolution and
 somehow slows down the evolution.

 In order to see the influence of the angular momentum to the
 gauge
 dynamics, let us analyze the canonical formulation of the DBI
 action more carefully. Now the action is
  \be
  S_p=-\tau_p\int dt
\frac{1}{\sqrt{H}}\sqrt{Det(h)(1-H(\dot{R}^2+R^2\dot{\theta}^2)-E^+_iD_{ik}E^-_k},
\ee with
 \bea
 E^{\pm}_i=E_i\pm
H(\dot{R}\partial_i{R}+R^2\dot{\theta}\p_i\theta)\\
h_{ij}=\delta_{ij}+F_{ij}+H(\partial_iR\partial_jR+R^2\p_i\theta\p_j\theta)\\
D_{ij}=Det(h)h^{-1}_{ij}. \eea

The canonical momenta conjugate to $R$ and $\theta$ are
 \bea
 \mathbf{\Pi}_R&=&\frac{\tau_p}{\sqrt{H}\sqrt{Det(h)(1-H(\dot{R}^2+R^2\dot{\theta}^2)-E^+_iD_{ik}E^-_k}}
 (Det(h)H\dot{R}-\frac{E^+_kD_{ki}-D_{ik}E^-_k}{2}H\partial_iR)\\
 \mathbf{\Pi}_\theta&=&\frac{\tau_p}{\sqrt{H}\sqrt{Det(h)(1-H(\dot{R}^2+R^2\dot{\theta}^2)-E^+_iD_{ik}E^-_k}}
 (Det(h)HR^2\dot{\theta}-\frac{E^+_kD_{ki}-D_{ik}E^-_k}{2}HR^2\partial_i\theta).
 \eea
 And the conserved energy density and momentum density are
 \bea
 \mathcal{H}&=&\sqrt{|\vec{\mathbf{\Pi}}_e|^2+\frac{\mathbf{\Pi}_R^2}{H}+\frac{\mathbf{\Pi}_\theta^2}{HR^2}
 +H(\mathbf{\Pi}_e^i\partial_iR)^2+HR^2(\mathbf{\Pi}_e^i\partial_i\theta)^2+|\vec{\mathcal{P}}|^2+\frac{Det(h)}{H}}\\
\mathcal{P}_i&=&-F_{ik}\mathbf{\Pi}_e^k-\partial_iR\mathbf{\Pi}_R-\p_i\theta\mathbf{\Pi}_\theta.
\eea Let us introduce a new set of fields
 \bea
 \p_iT=\sqrt{H}\p_iR & &
 \hspace{5ex}\mathbf{\Pi}^T=\frac{\mathbf{\Pi}_R}{\sqrt{H}}\\
 \p_iS=\sqrt{H}R\p_i\theta& &
 \hspace{5ex}\mathbf{\Pi}^S=\frac{\mathbf{\Pi}_\theta}{\sqrt{H}R}.
 \eea
The field $T$ could be identified with the tachyon field and the
field $S$ differs from $\theta$ by a constant rescaling in the
near horizon region. Similar to the case without angular momentum,
one could take $T$ and $S$ as the new components of gauge field
along two imagined  directions:
 \be
 A_T=T, \hspace{5ex}A_S=S.
 \ee
The Hamiltonian near the tachyon vacuum takes the same form as
(\ref{end}) but now $I,J=1,2,\cdots, p$ and $T,S$. Therefore, the
study of the fluctuations around the tachyon vacuum is very
similar to the one in the case without angular momentum. The only
difference is that apart from the fluctuations of the geometric
tachyon direction $R$, one has also the angular fluctuations.
Similarly, one may treat the Hamiltonian dynamics of the system
with angular momentum around the vacuum as the fluid equations.
The fluid equations look like (\ref{fluid}), and the integrability
condition is now
 \be
 v^i=-F_{ij}n^j-n^T\p_i T-n^S\p_i S, \hspace{3ex} v^T=\p_iTn^i,
 \hspace{3ex} v^S=\p_i S n^i
 \ee
 where
 \be
 n^S=\frac{\mathbf{\Pi}^S}{\mathcal{H}}, \hspace{5ex}
 v^S=\frac{\p_iS\mathbf{\Pi}^i}{\mathcal{H}}.
 \ee
The static solutions describe the various distributions of the
tachyon matter, string fluid and angular momentum. The explicit
solutions can be obtained straightforwardly. If one allows
momentum density along a direction, say $x^1$, the evolution of
the tachyon could not be completely homogeneous. Let us turn off
the magnetic field and boost along $x^1$. In this case, one has
 \be
 v^1=-\frac{\mathbf{\Pi}_R}{\mathcal{H}}\p_iR-\frac{\mathbf{\Pi}_\theta}{\mathcal{H}}\p_i\theta.
 \ee
Near the vacuum, $\mathbf{\Pi}_R\propto \sqrt{H}$ so the
inhomogeneous evolution of the radial direction is greatly
suppressed, as in the case without angular momentum. On the other
hand, $\p_i\theta$ could be finite.

Effectively, in this geometric tachyon setup, one may combine $R$
and $\theta$ into a complex tachyon field. However due to the
geometric nature, there exist a conserved angular momentum, which
changes the dynamics a little bit.

\section{The Radial Dynamics of (Dp, NS5)' System}

Now let us turn to another interesting case when one of the
transverse direction of NS5-Dp system is compactified on a circle
of radius $r_0$. We parameterize the compactified dimension by
$y$, and the three uncompactified dimensions by $r$, $\theta$,
$\phi$.

Now the metric is
\begin{eqnarray}\label{metric}
ds^2&=&dx^\mu dx_\mu+H(r)dx^ndx_n\equiv g_{MN}dx^Mdx^N\nonumber\\
\frac{g_s(\Phi)}{g_s}&=&\exp(\Phi-\Phi_0)=\sqrt{H(r)}\nonumber\\
H(r)&=&1+\frac{kl_s^2
\sinh\frac{r}{r_0}}{2rr_0(\cosh\frac{r}{r_0}-\cos\frac{y}{r_0})},
\end{eqnarray}
We are interested in the homogeneous evolution, so the action is
\begin{eqnarray}
S&=&-\tau_pV\int dt
\frac{1}{\sqrt{H(y,r)}}\sqrt{1-H(y,r)(\dot{y}^2+\dot{\vec{r}}^2)}
=-\tau_pV\int dt \mathcal{L}\nonumber\\
\mathcal{L}&=&\sqrt{\frac{1}{H}-\dot{y}^2-\dot{r}^2-r^2\dot{\theta}^2}
\end{eqnarray}
Therefore, one may take $H^{-1}$ as the effective potential for
the scalar fields. It is not hard to find out that $r=0, y=2n\pi
r_0$ is at the minima of the potential and is stable, while $r=0,
y=(2n+1)\pi r_0$ is a saddle point, where $r=0$ is stable but
$y=(2n+1)\pi r_0$ is unstable with tachyonic fluctuations. In
\cite{Kutasov2}, it was argued that when the Dp-brane stay at the
saddle point, it behaves as an unstable non-BPS brane and there
exist a kink solution which makes (Dp, NS5)'-system BPS after
tachyon condensation. More precisely, one of the Dp-brane
worldvolume direction changes to the compactifed $y$ direction so
that the overlap dimensions between Dp-brane and NS5-branes are
$p$ rather than $p+1$. This is a very interesting observation and
shed new light on the study of non-BPS brane. But actually, the
tachyon condensation could have many other channels besides
forming kink solution. It is very possible that the Dp-brane still
falls to the NS5-branes from $y=\pi r_0$ to $y=0$. Also even we
fix at $y$ direction, the radial behavior of Dp-brane in the three
noncompactified directions is quite interesting in its own right.
In this section, we would like to address this issue from the
analysis of DBI action.

Note that for simplicity, we turn off all the electromagnetic
fields. When we discuss the homogeneous evolution, the electric
field is always a constant. It is straightforward to take into
account the contribution of the electric field, as we have done in
the last section.

Before we turn to the analysis of the tachyon condensation without
forming kink, we would like to point out that there is an issue on
the gauge dynamics of the kink solution. In \cite{Sen3,Sen5}, it
has been shown how to construct the fundamental strings ending on
the BPS kink solutions in the nonBPS brane effective action. After
field redefinition, it is easy to analyze the same issue in the
gauge dynamics of kink solution of (Dp, NS5)'-system.

 The density of canonical momentum and Hamiltonian are
\begin{eqnarray}\label{momentum}
\mathbf{\Pi}_\theta=\frac{\tau_p r^2\dot{\theta}}{\mathcal{L}},& &\mathbf{H}=\frac{\tau_p}{\mathcal{L}H}\nonumber\\
\mathbf{\Pi}_y=\frac{\tau_p\dot{y}}{\mathcal{L}},& &
\mathbf{\Pi}_r=\frac{\tau_p\dot{r}}{\mathcal{L}}
\end{eqnarray}
and the equations of motions read as:
\begin{eqnarray}\label{comeom}
\mathbf{\Pi}_\theta&=&L\nonumber\\
\mathbf{H}&=&E\nonumber\\
\mathbf{\dot{\Pi}_y}&=&-\frac{\tau_p
kl_s^2\sinh\frac{r}{r_0}\sin\frac{y}{r_0}}{4\mathcal{L}H^2r_0^2r(\cosh\frac{r}{r_0}-\cos\frac{y}{r_0})}\\
\mathbf{\dot{\Pi}_r}&=&\frac{\tau_p}{2\mathcal{L}}\left(\frac{kl_s^2}{2H^2r_0r(\cosh\frac{r}{r_0}-\cos\frac{y}{r_0})^2}
\left(-\frac{\sinh\frac{r}{r_0}(\cosh\frac{r}{r_0}-\cos\frac{y}{r_0})}{r}+\frac{1-\cos\frac{y}{r_0}\cosh\frac{r}{r_0}}{r_0}\right)
+\frac{2L^2}{r^3E^2H^2}\right) \nonumber
\end{eqnarray}
where $L$ and $E$ are conserved angular momentum and energy
respectively.

It is also instructive to analyze the strength tensor. From
\cite{Chen1}, we have
\begin{eqnarray}
T^{\mu\nu}=-\tau_p\frac{\sqrt{-detG}}{\sqrt{H}}\left(\begin{array}{cccc}
\frac{1}{det(G)}&0&0&0\\
0&1&0&0\\
0&0&1&0\\
0&0&0&1\\
\end{array}\right)=-\tau_p\mathcal{L}\left(\begin{array}{cccc}
\frac{-1}{H\mathcal{L}^2}&0&0&0\\
0&1&0&0\\
0&0&1&0\\
0&0&0&1\\
\end{array}\right)
\end{eqnarray}
So the pressure is
\begin{eqnarray}
p=-\tau_p\mathcal{L}=-\frac{\tau_p}{EH}
\end{eqnarray}

Generically,  the equations of motions (\ref{comeom}) are too
involved to be solved exactly. Nevertheless, there are several
cases that one can deal with to get a qualitative picture of the
dynamics. It's clear from (\ref{comeom}) that when we set $y=n\pi
r_0,\dot{y}=0$ as initial condition, the Dbrane will not evolve in
the $y$ direction, no matter how it evolves in the other
transverse directions. Firstly we fix $y=\pi r_0$  and the
equation of motion of $r$ can be cast into an effective
one-dimensional problem:
\begin{eqnarray}\label{veff}
\dot{r}^2=-V_{eff}=\frac{1}{H}-\frac{1}{E^2H^2}(\tau_p^2+\frac{L^2}{r^2})
\end{eqnarray}
and the equation on $\theta$ is
 \be
 \dot{\theta}=\frac{L}{r^2EH}
 \ee

 In practice, one can learn the qualitative feature of the radial dynamics from the effective
potential. It's illuminating to see its behavior at large and
small $r$, compared to $r_0$. Notice that the minimal length of
$r_0$ cannot be smaller than the string scale $l_s$.

When $r/r_0$ is large, up to order $\frac{l_s}{r}$,
$H=1+\frac{kl_s^2}{2r_0r}$, and
\begin{eqnarray}
V_{eff}=(\frac{\tau_p^2}{E^2}-1)+\frac{kl_s^2}{2r_0r}(1-\frac{2\tau_p^2}{E^2})+\frac{L^2}{E^2r^2}+O(\frac{1}{r^3})
\end{eqnarray}
where we omit the terms proportional to
$(\frac{kl_s^2}{2r_0r})^2$.

When $r/r_0$ is small,
$H=1+\frac{kl_s^2}{4r_0^2}-\frac{kl_s^2r^2}{12r_0^4}+O(k(\frac{l_s}{r_0})^2(\frac{r}{r_0})^{2+\varepsilon})\,\,\,\,\,(\varepsilon>0)$.
After neglecting square and higher order of $(\frac{r}{r_0})^2$,
one can take $H$ as a constant $H_0=1+\frac{kl_s^2}{4r_0^2}$ and
\begin{eqnarray}
V_{eff}(r)=(\frac{\tau_p^2}{E^2}-1)+\frac{L^2}{E^2r^2}
\end{eqnarray}

Therefore we see that in both cases, the effective potential is
well approximated by
\begin{eqnarray}\label{effpi}
V_{eff}(u)=au^2+bu+c
\end{eqnarray}
 with $ u=\frac{1}{r}$ and
 \begin{eqnarray}
 a=\frac{L^2}{E^2}
,\,\,\,b=\frac{kl_s^2}{2r_0}(1-2\frac{\tau_p^2}{E^2}),\,\,\,
c=\frac{\tau_p^2}{E^2}-1,\hspace{3ex} \mbox{for large $r/r_0$} \nonumber\\
a=\frac{L^2}{E^2H^2_0},\,\,\,
b=0,\,\,\,c=\frac{\tau_p^2}{E^2H^2_0}-\frac{1}{H_0},\hspace{3ex}
\mbox{for small $r/r_0$}
\end{eqnarray}
Depending on the parameters, the equation $V_{eff}(u)=0$ could
have two different roots $u_1 < u_2$, one root $u_0$ or no root.
In fact, the one-dimensional problem we have is quite similar to
the radial motion of an object in a Newtonian gravitational
system. Remarkably,  this effective potential captures the
qualitative feature of the exact effective potential in
(\ref{veff}). In Fig. 1, through numerical analysis, we draw off
the relation between $V_{eff}$ and $r$, by adjusting the conserved
quantity $E$ and $L$. From the form of the potential, one can read
out the possible radial motions of the D-brane. We will see that
the four possibilities could be compared with the
semi-quantitative analysis in the large $r/r_0$ region.

Let us discuss the radial behavior case by case. When $r/r_0$ is large, we have
\begin{enumerate}
\item[(1)] when the angular momentum $L=0$, if $E< \tau_p$, the
D-brane moves in a restricted region $0<r< r_m$ with $r_m=1/u_0$;
otherwise it could escape to infinity;

\item[(2)] when $L\neq 0$,
and $b^2-4ac>0$,  $c>0$, there is an elliptic orbit, and the
pressure $p\sim\frac{1}{H}$ is a monotonically increasing function
of $r$, but varies slowly. In terms of $E, L$, we require $E<
\tau_p$ and
 \be \label{LE1}
\frac{L^2}{E^2} < \frac{(1+2c)^2}{4c}(\frac{kl_s^2}{2r_0})^2, \ee
such that the D-brane move between $r_{min}$ and $r_{max}$, where
$r_{min,max}$ are the inverse of roots of $V_{eff}=0$.

\item[(3)] when $L\neq 0$, and $b^2-4ac>0$, $c<0$, namely $E> \tau_p$,
Dbrane is moving in the range $(r_{min},\infty)$, where
$r_{min}=1/u_2$ is the inverse of the positive root of
(\ref{effpi}). The pressure $p$ increase as $r$ increase, getting
to asymptotically $\frac{\tau_p}{E}$ when $r$ large.

\item[(4)] when $L\neq 0$, and $b^2-4ac=0$, namely $E<\tau_p$ and
 \be \label{LE2}
\frac{L^2}{E^2} = \frac{(1+2c)^2}{4c}(\frac{kl_s^2}{2r_0})^2, \ee
 there is an circle orbit, the pressure is constant. The radius of
 the circle is
 \be
 r_c=\frac{1+2c}{2c}\frac{kl_s^2}{2r_0},
 \ee
so that in order to have circular orbit in the large $r/r_0$
region, one needs to require $k$ be very large or $E\simeq
\tau_p$;

\item[(5)] $b^2-4ac<0$ is not physical, actually it is
inconsistent with (\ref{momentum}).
\end{enumerate}
 In short, when $E< \tau_p$, the D-brane could not escape
to infinity and move in a restricted region. In particular, the
angular momentum play its role in a subtle way: when $L$ satisfies
(\ref{LE2}) there exist a stable circular orbit; when $L$
satisfies (\ref{LE1}), there exist an elliptic orbit; otherwise,
there is no motion at all. Qualitatively, the case (2)-(5) could
be mapped to four cases in Fig 1..

On the other hand, we are more interested in the radial behavior
of the D-brane when $r$ near its stable point, i.e. $r/r_0 \sim
0$. In this case, the pressure is a constant and
 \begin{enumerate}
 \item[(1)] when $L=0$, $E> \frac{\tau_p}{\sqrt{H_0}}$, it is of linear
motion with constant velocity, and could escape to infinity, while
$E\leq \frac{\tau_p}{\sqrt{H_0}}$ is inconsistent with
(\ref{momentum}), actually it is obvious from (\ref{momentum})
that $E\geq\frac{\tau_p}{\sqrt{H_0}}$.

\item[(2)] when $L\neq 0$, $E> \frac{\tau_p}{\sqrt{H_0}}$, it
could move between $1/u_2$ and $\infty$, $E\leq
\frac{\tau_p}{\sqrt{H_0}}$ is unphysical as stated above.
\end{enumerate}
Now the evolution of the Dp-brane is very different from the
uncompactified case. First notice that one has a limit on the
energy $E\geq\frac{\tau_p}{\sqrt{H_0}}$. More interestingly, we
have $r > \frac{L}{E\sqrt{H_0}}$, which means that due to the
existence of the angular momentum, the D-brane cannot reach $r=0$.

In both regions, we have $r^2\dot{\theta}\simeq L$. So the
evolution is much the same as that of a particle in Newtonian
gravitational field.
\begin{figure}
\includegraphics[width=0.24\textwidth, height=0.2\textheight]{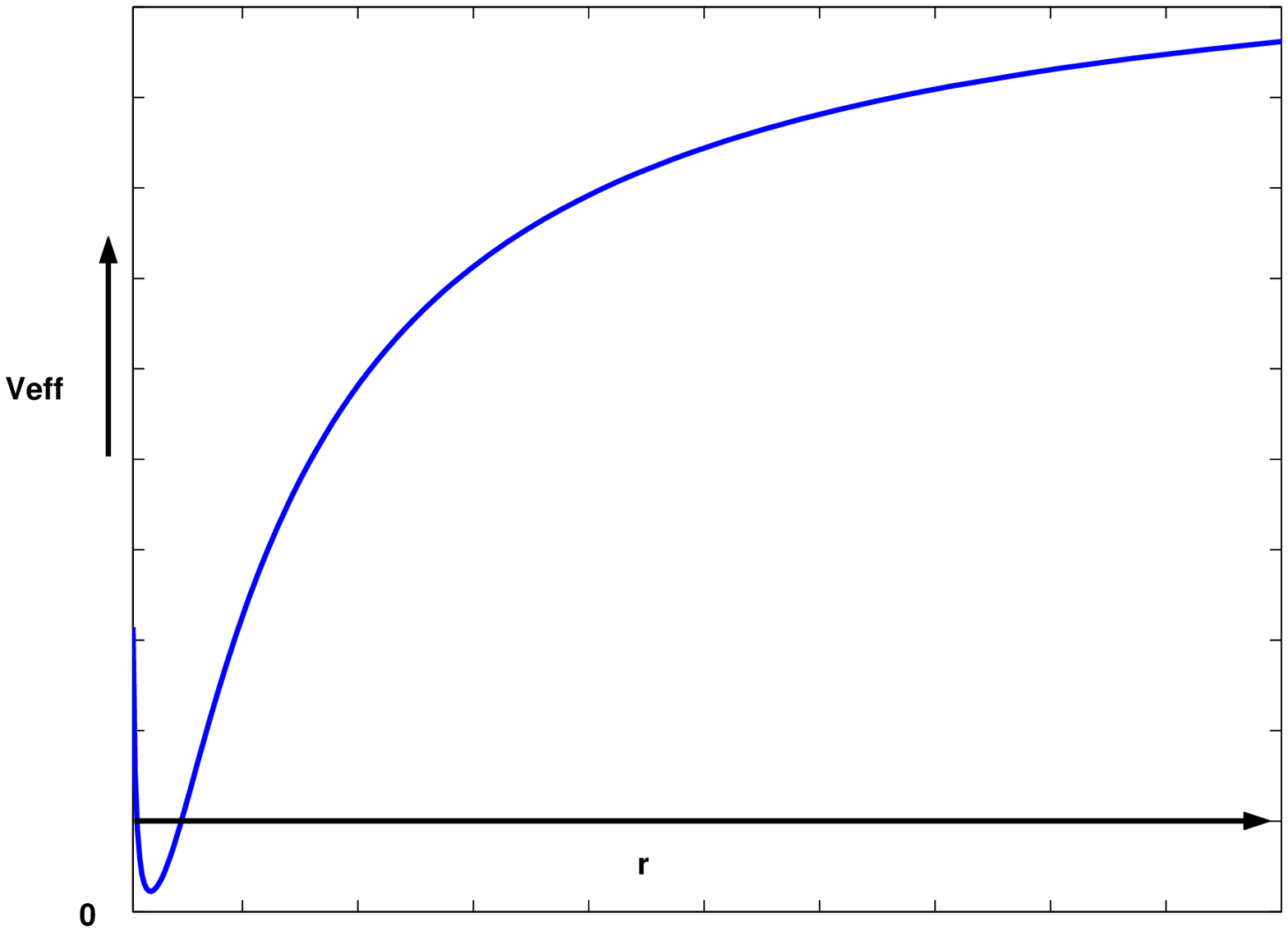}
\includegraphics[width=0.24\textwidth, height=0.2\textheight]{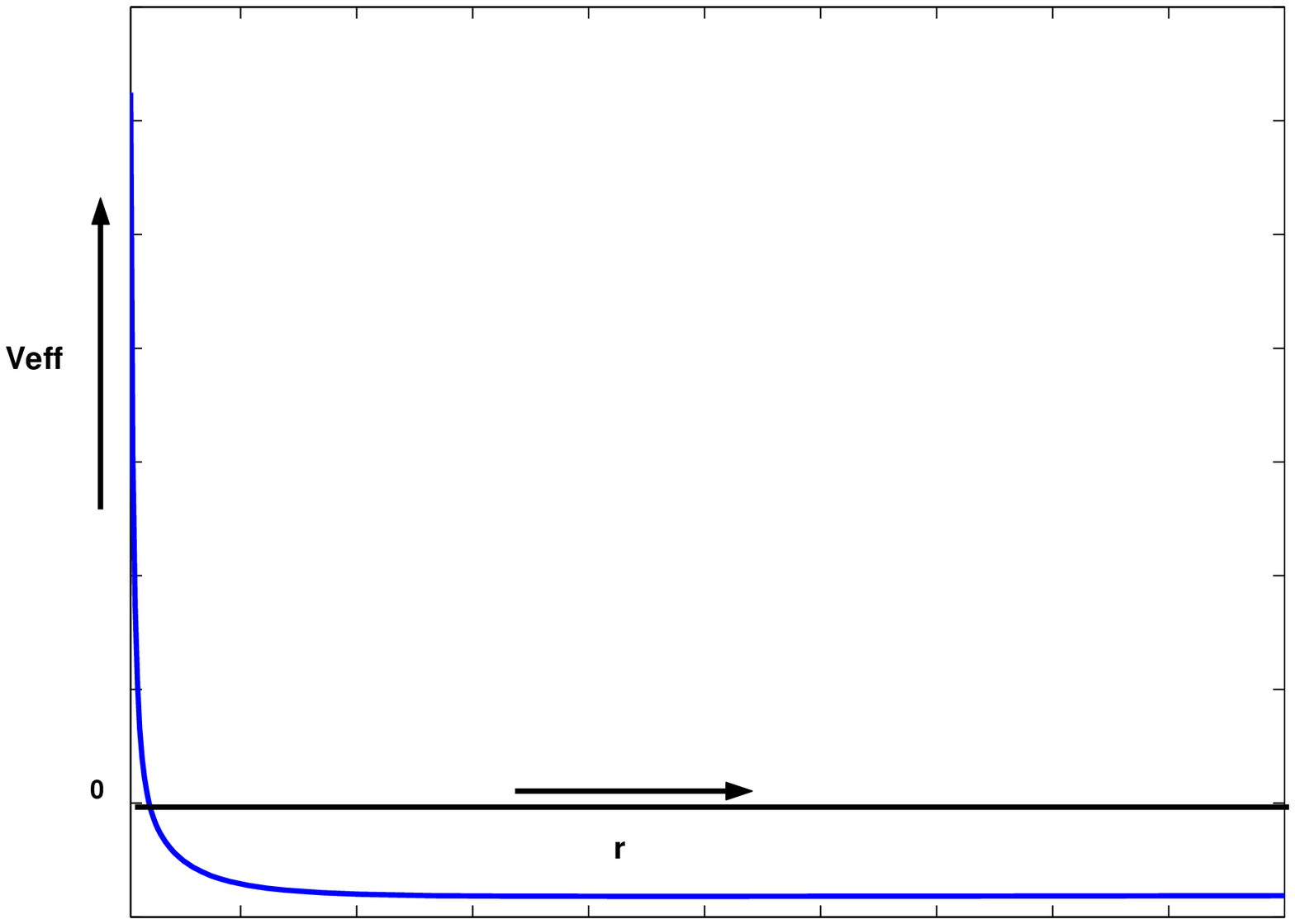}
\includegraphics[width=0.24\textwidth, height=0.2\textheight]{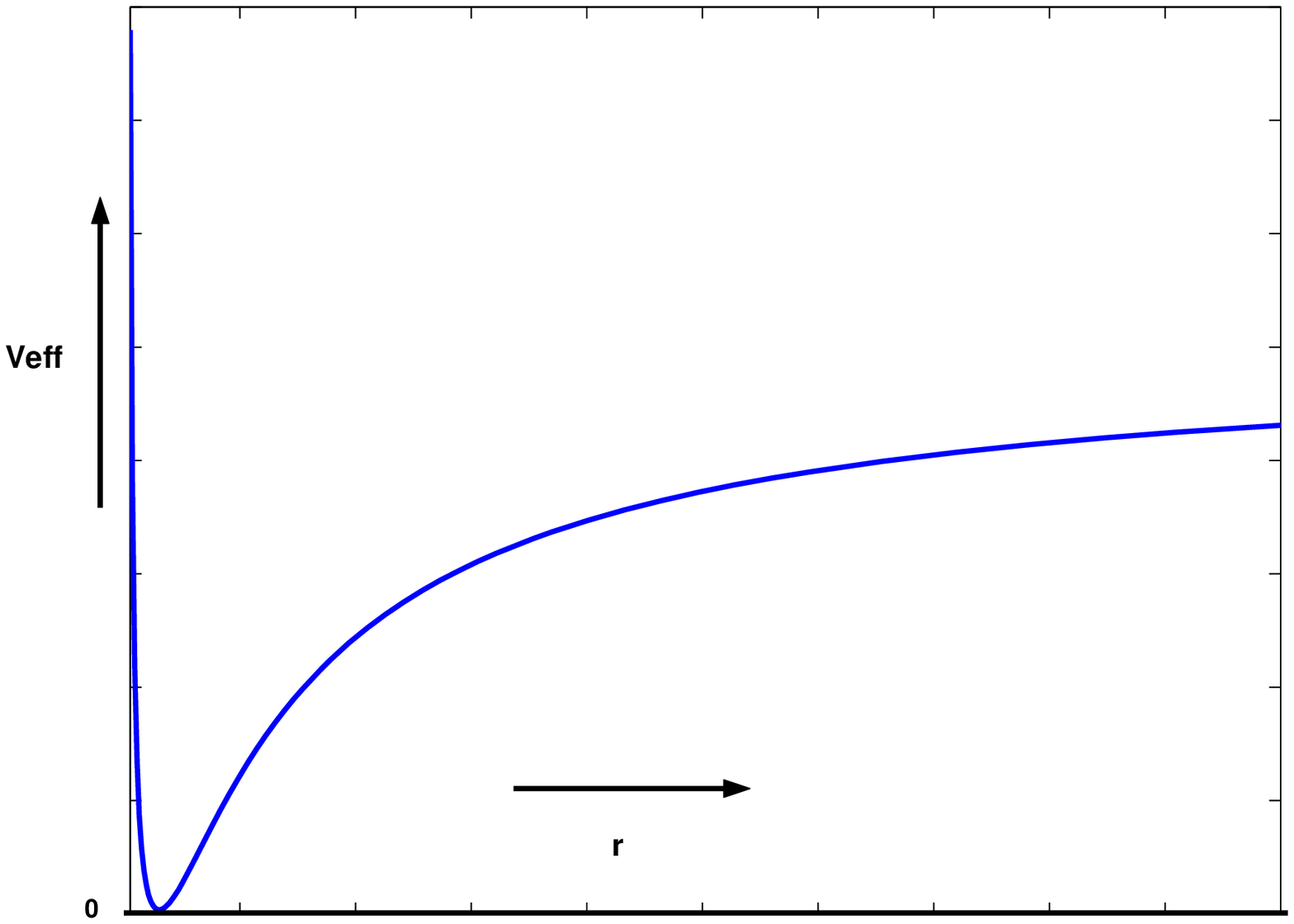}
\includegraphics[width=0.24\textwidth, height=0.2\textheight]{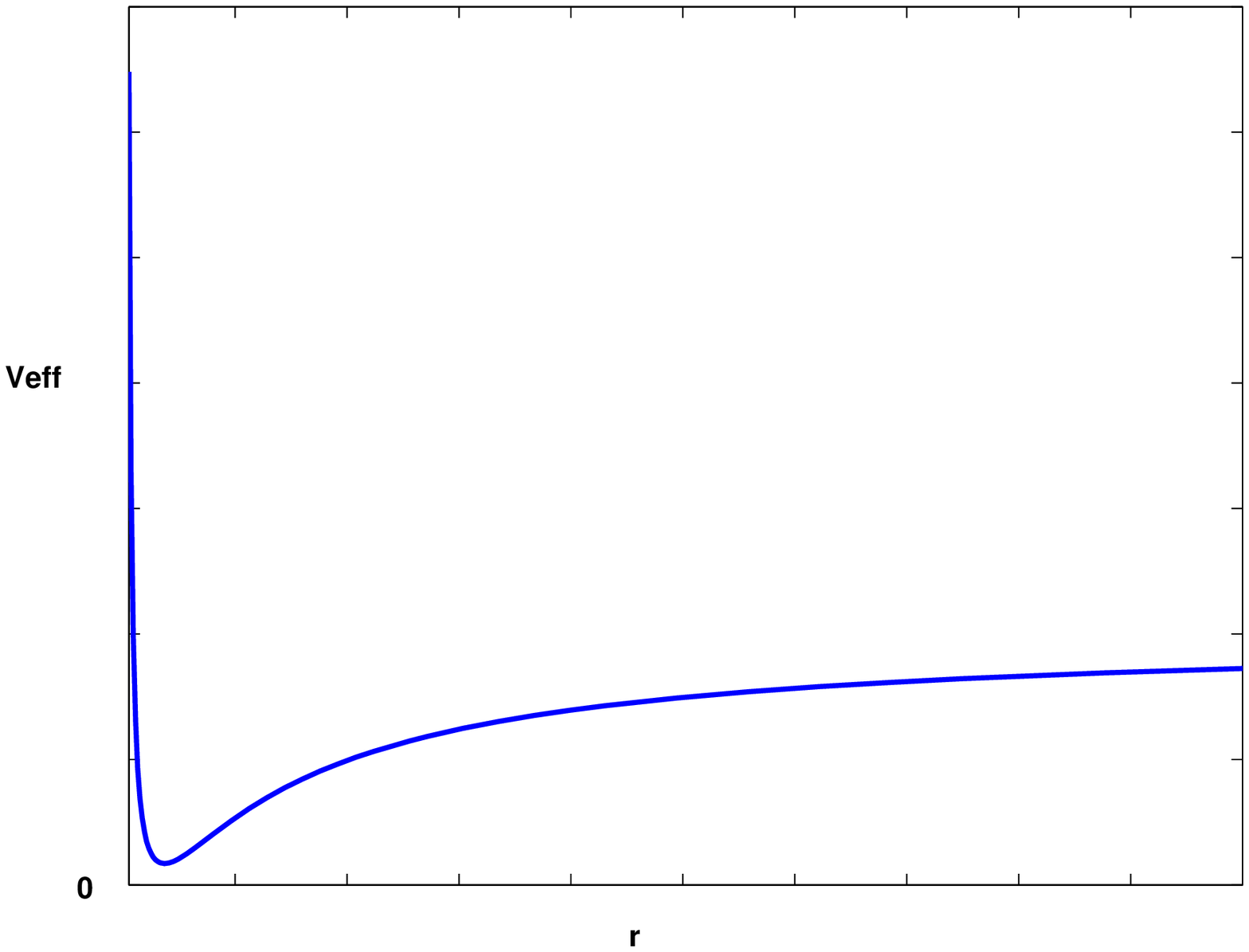}
\caption{y fixed at $\pi r_0$, case(2)-(5). These figures are
obtained by numerical analysis from the original potential
(\ref{veff}), and is semi-quantitatively same as (\ref{effpi}).}
\end{figure}

Let's try to combine the discussions in the two regions together
to have a picture of the Dp-brane radial motion: when $L=0$ and
$\tau_p>E> \frac{\tau_p}{\sqrt{H_0}}$, the Dp-brane could move in
a restricted region and get to $r=0$; when $L\neq 0$ and
$\tau_p>E> \frac{\tau_p}{\sqrt{H_0}}$, the D-brane may still move
in a restricted region, but could never reach $r=0$; when $E>
\tau_p$, the D-brane can escape to infinity.

If we instead fix $y=2n\pi r_0$, the large $\frac{r}{r_0}$
behavior is the same as above, but for small $\frac{r}{r_0}$,
\begin{eqnarray}
H&=&1+\frac{kl_s^2}{2r_0r}\frac{\sinh\frac{r}{r_0}}{\cosh\frac{r}{r_0}-1}\nonumber\\
  &=&1+\frac{kl_s^2}{r^2}+O(k(\frac{l_s}{r_0})^2(\frac{r}{r_0})^{2+\varepsilon})\,\,\,\,\,(\varepsilon>0)\nonumber\\
 &\simeq&\frac{kl_s^2}{r^2}
\end{eqnarray}
where in the last step, we use the near throat region
approximation. Obviously, it looks the same as the one in
uncompactified case\cite{Kutasov1}. Again we obtain approximate
effective potential that is semi-quantitatively consistent with
numerical analysis. Thus
\begin{eqnarray}\label{veff2}
V_{eff}=\frac{\tau_p^2}{k^2l_s^4E^2}r^4+\frac{L^2}{k^2l_s^4E^2}r^2-\frac{1}{kl_s^2}r^2
\end{eqnarray}

In this case we have if \begin{enumerate}
\item[(1)] $\frac{L}{E}<\sqrt{k}l_s$ and $E<\tau_p$, the
Dbrane is bounded near NS5 branes in the range $(0,r_{max})$, and
when it get close to NS5 branes, the pressure decrease as $r^2$;

\item[(2)] $\frac{L}{E}>\sqrt{k}l_s$ and $E<\tau_p$, this is
forbidden in the near throat region;

\item[(3)] $\frac{L}{E}<\sqrt{k}l_s$ and $E>\tau_p$, the Dbrane is
moving in the range $(0,\infty)$, the pressure behaves as
$\frac{\tau_pr^2}{Ekl_s^2}$ and increases very slowly as $r$
becoming larger, getting to asymptotic value
$p_{\infty}=\frac{\tau_p}{E}$;

\item[(4)] $\frac{L}{E}>\sqrt{k}l_s$ and $E>\tau_p$, there is a potential
barrier near NS5 brane, so the Dbrane will be scattered and can
never reach the near throat region. The pressure increases very
slowly as $r$ becoming larger, getting to asymptotic value
$p_{\infty}=\frac{\tau_p}{E}$.
\end{enumerate}
In the near throat region, one has to require $L<\sqrt{k}l_sE$. In
other words, if $L\geq \sqrt{k}l_sE$, the D-brane cannot enter the
near throat region. Actually, one can solve the equation exactly
and get the small $r/r_0$ behavior like (\ref{cosh}) but now
 \be
\alpha^2=\frac{1}{kl_s^2}(1-\frac{L^2}{kl_s^2E^2})\,\,,\,\,\,\,\beta^2=\frac{kl_s^2E^2}{\tau_p^2}(1-\frac{L^2}{kl_s^2E^2}).
\ee

We see that $r$ and pressure decrease exponentially.
 For $y$ fixed at $2n\pi r_0$, when $r$ is large, we
still have $\dot{\theta}\simeq\frac{L}{r^2E}$, but when $r$ is
small $\dot{\theta}\simeq\frac{L}{kl_s^2E}$, that is
$\theta\propto t$, keeping the angular momentum conserved.
\begin{figure}
\includegraphics[width=0.24\textwidth, height=0.2\textheight]{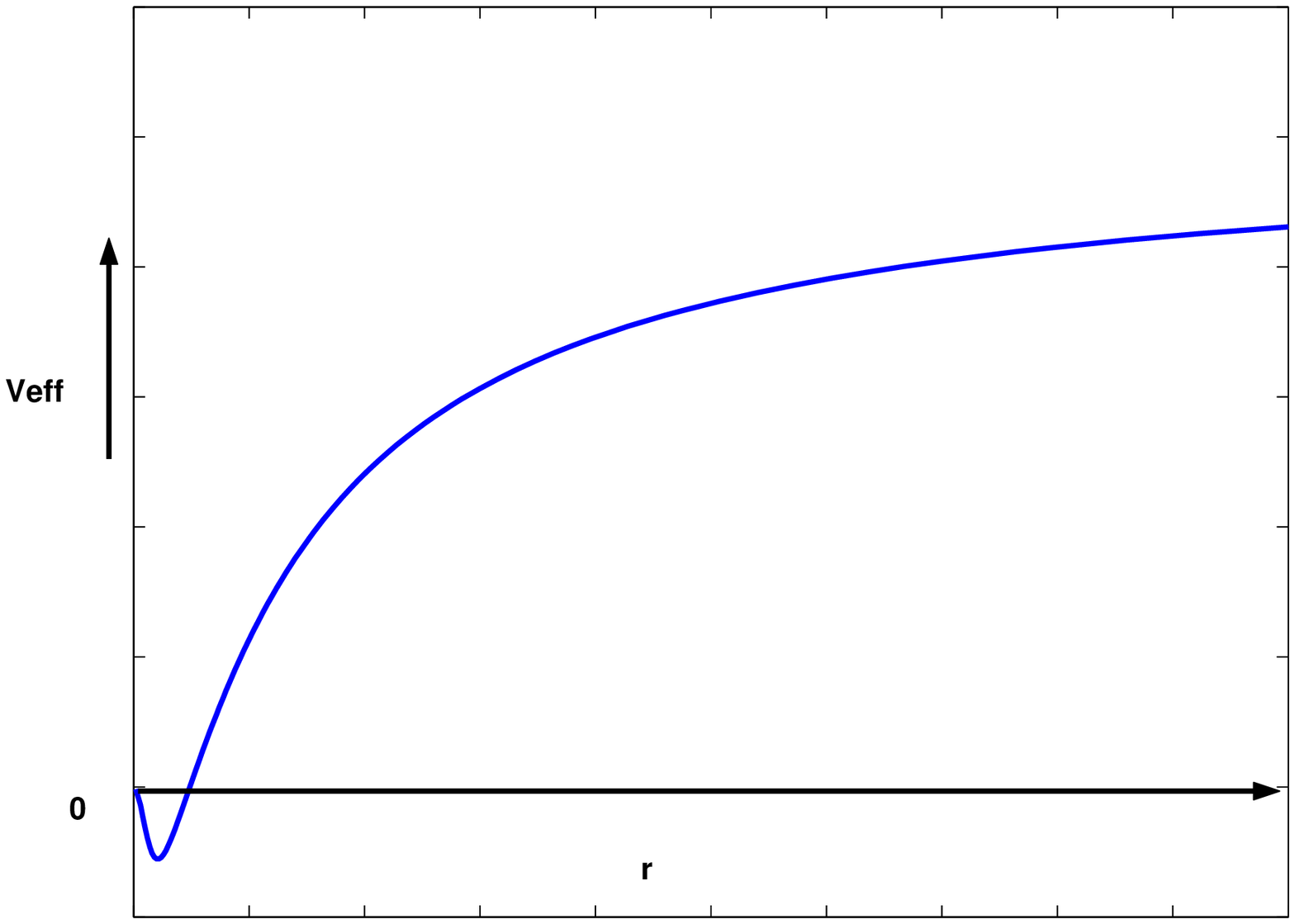}
\includegraphics[width=0.24\textwidth, height=0.2\textheight]{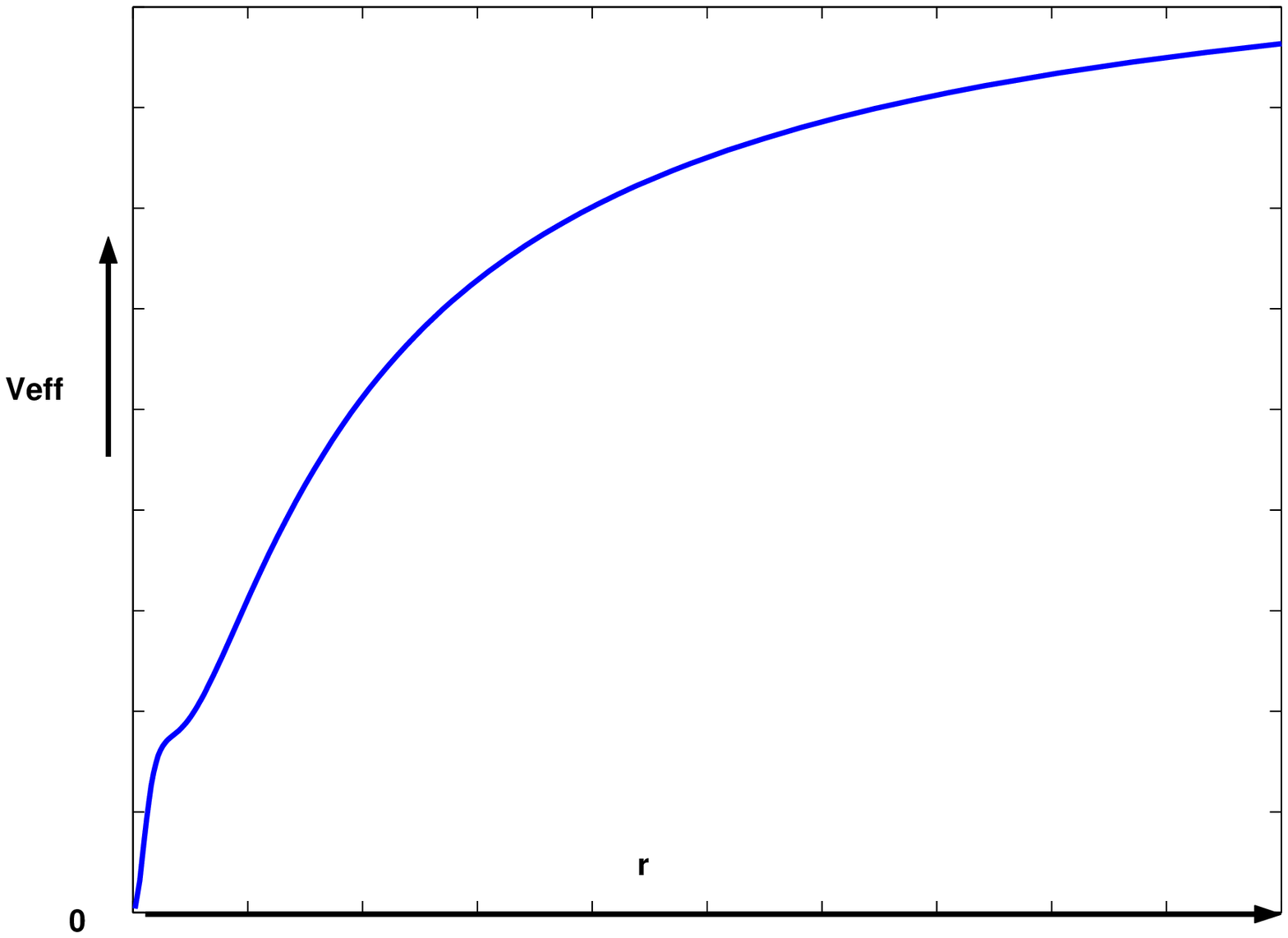}
\includegraphics[width=0.24\textwidth, height=0.2\textheight]{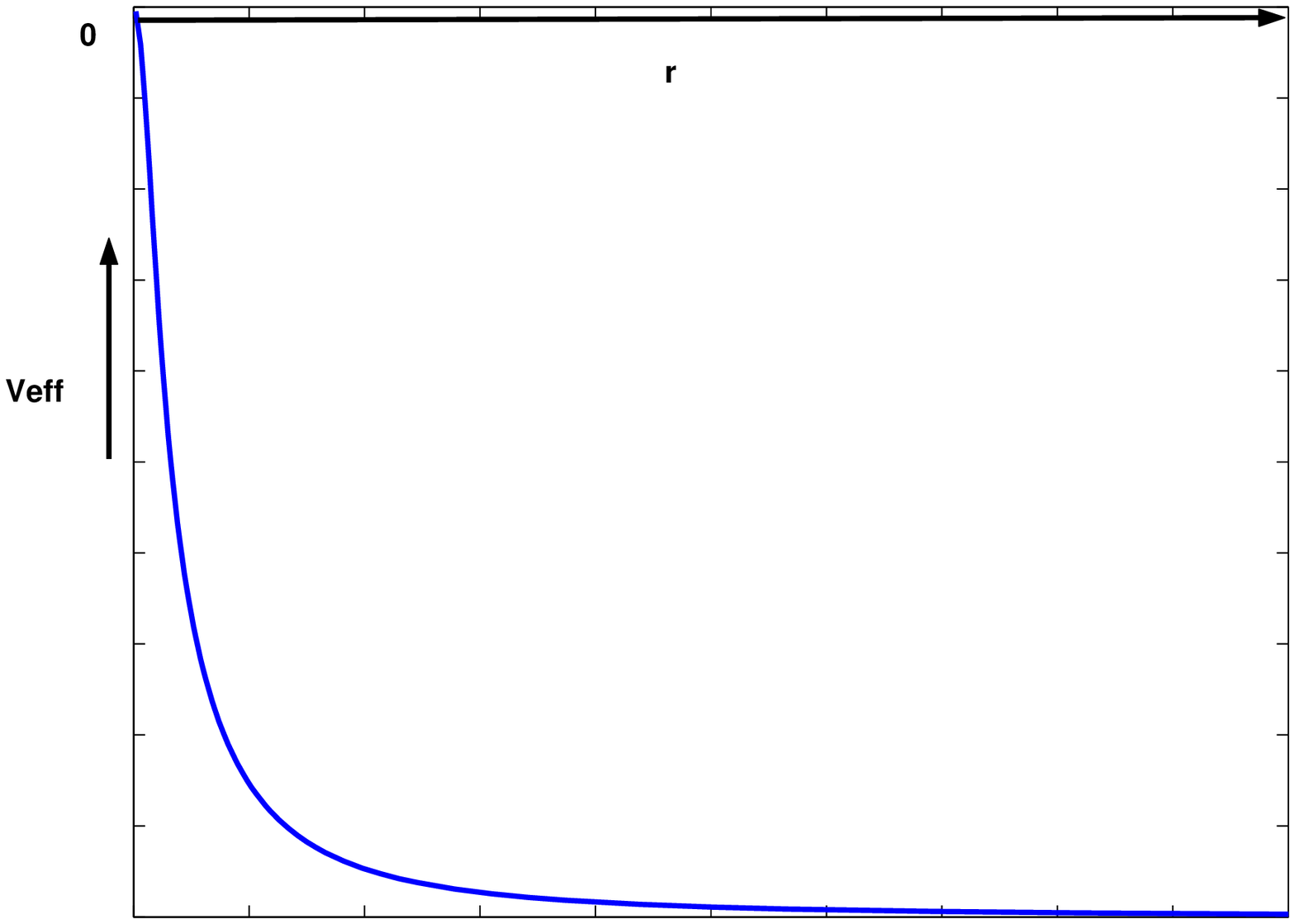}
\includegraphics[width=0.24\textwidth, height=0.2\textheight]{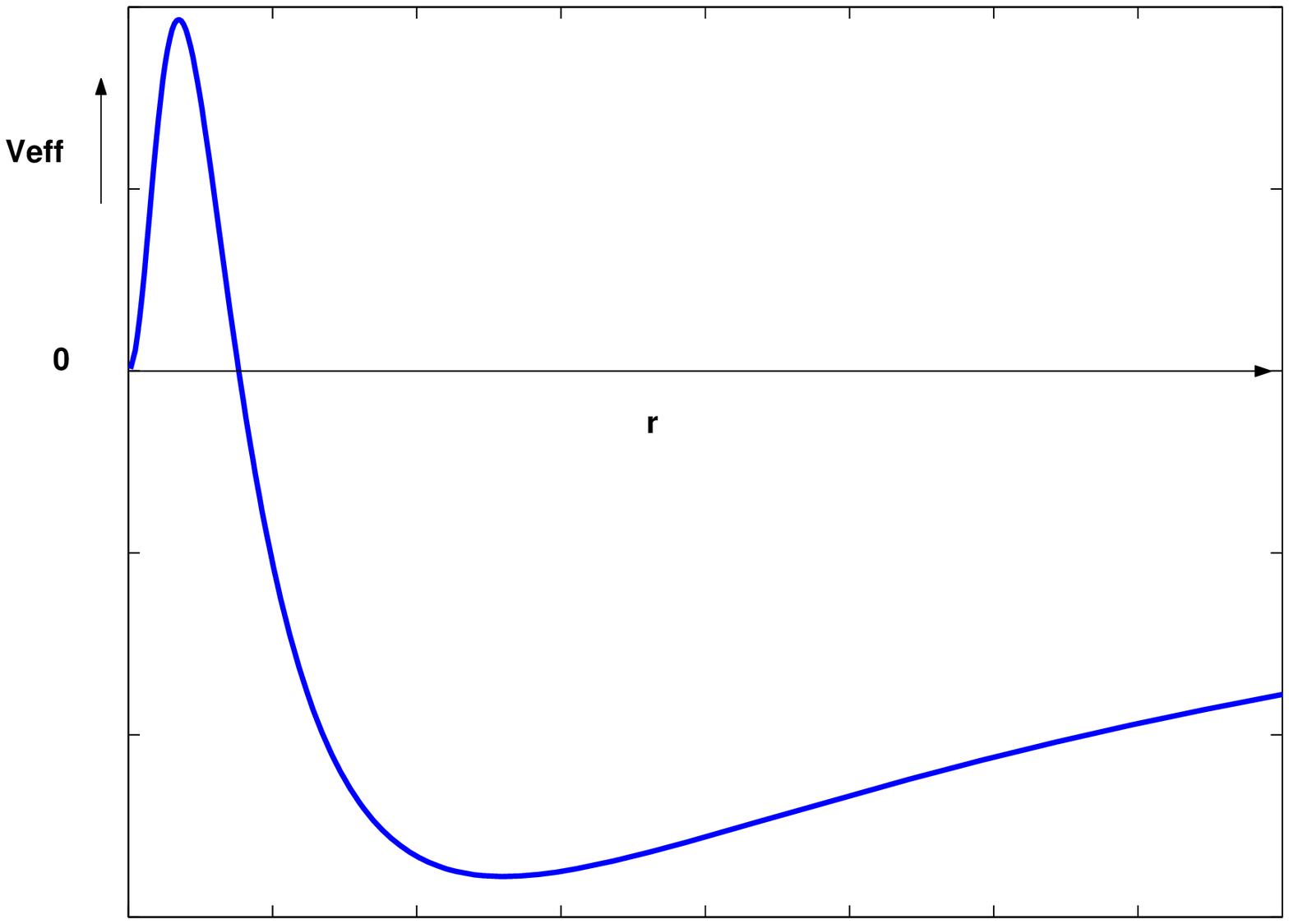}
\caption{y fixed at $2n\pi r_0$, case(1)-(4). These figures are
obtained by numerical analysis from the original potential
(\ref{veff}), and is semi-quantitatively same as (\ref{veff2}).}
\end{figure}


 Now let's consider more generally initial conditions.
From (\ref{comeom}), we see that when $0<y<\pi r_0$,
$\mathbf{\dot{\Pi}_y}<0$ and when $\pi r_0<y<2\pi r_0$,
$\mathbf{\dot{\Pi}_y}>0$. This is nothing but the fact that $y=\pi
r_0$ is at the peek of the potential and $y$ tends to roll down to
$y=0$.

For the $r$ direction,  the behavior is even more complicated.
When $y=\pi r_0$ we have
\begin{eqnarray}
\dot{\mathbf{\Pi}}_r=\frac{\tau_p}{2\mathcal{L}H^2}(\frac{kl_s^2}{2rr_0(\cosh\frac{r}{r_0}+1)}(-\frac{\sinh\frac{r}{r_0}}{r}+\frac{1}{r_0})+\frac{2L^2}{r^3E^2})
\end{eqnarray}
It is clear that in this case, when $L=0$,
$\dot{\mathbf{\Pi}}_r<0$, or say r=0 is the minimum of potential.
But nonvanishing angular momentum causes centrifugal force which
dominates when r is small($\sim\frac{1}{r^3}$). And when y is near
the unstable saddle point $\pi r_0$, D-brane feels repulsive force
in near throat region, and attractive force in far away region.
This fact is consistent with our argument that when $y=\pi r_0$,
the nonvanishing angular momentum forbids the Dp-brane from
approaching $r=0$.

When $y=2n\pi r_0$  we have
\begin{eqnarray}
\dot{\mathbf{\Pi}}_r=\frac{\tau_p}{2\mathcal{L}H^2}
\left(-\frac{kl_s^2}{2rr_0(\cosh\frac{r}{r_0}-1)}(\frac{\sinh\frac{r}{r_0}}{r}+\frac{1}{r_0})+\frac{2L^2}{r^3E^2}\right)
\end{eqnarray}
Again there is an attractive force and a repulsive centrifugal
force from angular momentum. The former dominate in the large
$r/r_0$ region, but in the near throat region the competition
depends on $\alpha=\frac{L^2}{kl_s^2E^2}$. Recall that one
requires $L< \sqrt{k}l_s E$ so that $\alpha <1$ and the total
force is still attractive.

Therefore, if we assume that the Dp-brane initially deviate from
$y=\pi r_0$ a little bit, it will roll down to the stable point
$y=0$. Let's see what is the radial motion of D-brane in the three
 uncompactified directions. We have
 \be
 \dot{y}^2+\dot{r}^2=-V_{eff}=\frac{1}{H}-\frac{1}{E^2H^2}(\tau_p^2+\frac{L^2}{r^2})
 \ee
 In the large $r/r_0$ region, $H=1+\frac{kl^2_s}{2r_0r}$ is
 independent of $y$. So as $y$ rolls down, the orbit of the radial
 motion changes if $\dot y$ is not a constant. But if $E<\tau_p$,
 the D-brane move in a restricted region and cannot escape to
 infinity. It could be possible that the D-brane move in an elliptic or
 circular orbit and never come near to the near throat region when $E\simeq \tau_p$. If
 the D-brane can move into the near throat region with $L<\sqrt{k}l_sE$, it may roll to
 $r=0$ since at $y=0$ the D-brane feel attractive force. However,
 the dynamics here is really complicated and depends on the
 initial conditions. For example, even when $y$ rolls to $y=0$ but
 with nonzero $\dot{y}$, the D-brane will move in a restricted region rather than reaching $r=0$ directly.
 As $y$ oscillate around $y=0$ and finally frozen to $y=0$, the
 D-brane will condensate to $y=0$ at the end.

The gauge dynamics around the tachyon vacuum is very similar to
the one without compact transverse dimension. At the end of the
tachyon potential, the Hamiltonian could be rewritten as
 \be
 \mathcal{H}=\sqrt{\mathbf{\Pi}^2_e+\frac{\mathbf{\Pi}^2_y}{H}+\frac{\mathbf{\Pi}^2_r}{H}+\frac{\mathbf{\Pi}_\theta^2}{r^2H}+\frac{\tau_p^2}{H}}.
 \ee
Obviously, the energy consists of the kinetic energy of $y$ and
$r$, the flux energy and the one from angular momentum. The gauge
dynamics could be studied in the same way as in last section. The
key point is to treat the field $y$ also as a tachyon field
component and the dynamics could still be cast into a fluid
equation. We will not repeat it here.

\section{Conclusions and Discussions}
In this paper, we investigated in details the radial motion of
Dp-brane near NS5-branes without and with one compactified
transverse direction, using the DBI effective action of (Dp,
NS5)-system. In the case without compactified dimension, we
studied the dynamics with the world volume gauge fluxes and the
angular momentum. We mainly focused on the
homogeneous evolution, in which case the electric fields on the
D-brane are conserved.
With various initial conditions, the D-brane could move in a
restricted region or escape to infinity. But there is no
stationary orbit. The influence of the gauge field on the dynamics
could be studied in the Hamiltonian formulation. At the end of the
tachyon condensation, the pressureless remnants has two
components, a string fluid and the tachyon matter. Around the
tachyon vacuum, the Hamiltonian dynamics of our geometric tachyon
system is very similar to the one in the rolling tachyon case,
after simple field redefinition. The dynamics could be described
by an effective fluid equations augmented by integrability
condition. Various issues in the rolling tachyon case could be
carried over to our geometric tachyon configuration. Even taking
into account of the angular momentum, the product of the tachyon
condensation is still pressureless. The angular momentum will slow
down the falling the Dp-brane and lowers the critical value of the
electric field.

In the case with one transverse dimension compactified, we also
study D-brane dynamics with various initial conditions, both in
the large and small radial distance regions. The dynamics is
different if we initially put the D-brane at the meta-stable
point($y=\pi r_0$, $r_0$ the compactification radius) or at the
stable point($y=2n\pi r_0$). The former case shows quite different
features from the uncompactified case. By appropriate
approximation, the radial motion is equivalent to a point particle
moving in a Newtonian potential. It is familiar that in this case
one can have elliptic orbit, circular orbit, and unbounded orbit
corresponding to different initial conditions. Due to the
existence of the angular momentum, the Dp-brane cannot reach
$r=0$.  In the latter case, the dynamics is very reminiscent of
that of uncompactified case, the radial coordinate as well as the
pressure falls exponentially with time in the near throat region.
If initially $y$ deviate from $y=\pi r_0$, it will rolls to $y=0$
and the radial motion of D-brane in the three uncompactified
directions depends on the initial conditions. If $E<\tau_p, L<
\sqrt{k}l_sE$, the radial motion could be stable orbit in the
large $r$ region, but more possibly it rolls to $r=0$ at the end.

 It might be very interesting to investigate what these
dynamics corresponds to in the Little String Theory. In
\cite{Sahakyan}, it was argued that the D-brane could be taken as
the defect in the dual LST. It is not clear how to describe the
dynamics here in the dual picture. In \cite{Sen4, Pil4}, in the
framework of the DBI effective action, the open/close duality has
been investigated in some details. It would be nice to understand
this issue in (NS5, D)-system. It also deserves further study if
we can have a stringy treatment beyond effective action of the
various dynamical processes.

As we have shown in detail, the radial dynamics of the (Dp
NS5)-systems are very similar to the rolling tachyon, but it also
has something new. The input of the angular momentum make the
dynamics of the (Dp NS5)-systems much richer. It would be
interesting to study the inhomogeneous evolution of the system as
have been done in the tachyon case\cite{Felder}.

\section*{Acknowledgements}

The work of BS was supported by a grant of NSFC and a grant of
Chinese Academy of Sciences, the work of BC was supported by NSFC
grant 10405028 and a grant of Chinese Academy of Sciences.

\bibliographystyle{amsplain}

\begin{thebibliography}{100}

\bibitem{Dai}J. Dai, R. G. Leigh and J. Polchinski, Mod. Phys.
Lett. {\bf A4} (1989)2073.\\
R. G. Leigh, \textit{Dirac-Born-Infeld action from Dirichlet Sigma
model}, Mod. Phys. Lett. {\bf A4} (1989)2767.
\bibitem{Sen1}A. Sen, \textit{Supersymmetric World-volume Action for Non-BPS
D-branes}, JHEP {\bf 9910}(1999)008. \\
M. R. Garousi, \textit{Tachyon couplings on non-BPS D-branes and
Dirac-Born-Infeld action}, Nucl.Phys. {\bf B584} (2000) 284-299,
hep-th/0003122.
\bibitem {Sen2} A. Sen, \textit{Tachyon matter}, JHEP \textbf{0207} (2002)065,
hep-th/0203265.\\
 A. Sen, \textit{Field theory of tachyon matter},
Mod. Phys. Lett. A \textbf{17} (2002)1797, hep-th/0204143.
\bibitem{Sen4}A. Sen, \textit{Open-closed duality at tree level},
Phys. Rev. Lett. {\bf 91} (2003)181601, hep-th/0306137.\\
A. Sen, \textit{Open-closed duality: lessons from matrix model},
Mod. Phys. Lett. A{\bf 19} (2004)841, hep-th/0308068.
\bibitem{Kutasov1}D. Kutasov, \textit{D-brane dynamics near
NS5-branes}, hep-th/0405058.
\bibitem{Kutasov2}D. Kutasov, \textit{A geometric interpretation of
the open string tachyon}, hep-th/0408073.
\bibitem{Nakayama}Y. Nakayama, Y. Sugawara and H. Takayanagi,
\textit{Boundary states for the rolling D-branes in NS5
background}, JHEP \textbf{0407} (2004)020, hep-th/0406173.
\bibitem{Sahakyan} D. Sahakyan, \textit{Comments on D-brane
dynamics near NS5-branes}, JHEP \textbf{0410}
(2004)008,hep-th/0408070.
\bibitem{Chen1} Bin Chen, Miao Li, Bo Sun, \textit{Dbrane Near NS5-branes: with Electromagnetic Field
}, JHEP {\bf 0412}(2004)037, hep-th/0412022.
\bibitem{Rey1}Y. Nakayama, K. L. Panigrahi, S. J. Rey and H.
Takayanagi, \textit{Rolling Down the Throat in NS5-brane
Background: The Case of Electrified D-Brane}, hep-th/0412038.
\bibitem{Rey2}D. Bak, S.J. Rey and H. U. Yee, \textit{Exactly Soluble Dynamics of
(p,q) String Near Macroscopic Fundamental Strings},
hep-th/0411099.\\
J. Kluson, \textit{Non-BPS Dp-Brane in the Background of
NS5-Branes on Transverse $R^3xS^1$}, hep-th/0411014. \\
S. Thomas and J. Ward, \textit{D-Brane Dynamics and NS5 Rings},
hep-th/0411130.\\
A. Ghodsi and A. E. Mosaffa, \textit{D-brane Dynamics in RR
Deformation of NS5-branes Background and Tachyon Cosmology},
hep-th/0408015.\\
K. L. Panigrahi, \textit{D-Brane Dynamics in Dp-Brane Background},
Phys. Lett. B{\bf 601}(2004)64, hep-th/0407134.\\
H. Yavartanoo, \textit{Cosmological Solution from D-brane motion
in NS5-Branes background}, hep-th/0407079.

\bibitem{Pil1}Gary Gibbons, Kentaro Hori, Piljin Yi, \textit{String Fluid from Unstable
D-branes}, Nucl.Phys. B596 (2001) 136-150, hep-th/0009061.
\bibitem{Pil2}G. Gibbons, K. Hashimoto and P. Yi, \textit{Tachyon
Condensates, Carrollian Contraction of Lorentz Gruop, and
Fundamental Strings}, JHEP {\bf 0209}(2002)061, hep-th/0209034.
\bibitem{Pil3}O-Kab Kwon and Piljin Yi, \textit{String Fluid, Tachyon Matter, and Domain
Walls}, JHEP {\bf 0309} (2003)003, hep-th/0305229.
\bibitem{Pil4}Ho-Ung Yee and Piljin Yi, \textit{Open/Closed Duality, Unstable D-Branes, and Coarse-Grained Closed
Strings}, Nucl.Phys. B{\bf 686} (2004)31-52, hep-th/0402027.
\bibitem{Sen3}A. Sen, \textit{Open and closed strings from
unstable D-branes}, Phys. Rev. D 68(2003)106003.
\bibitem{Sen5}A. Sen, \textit{Fundamental Strings in Open String Theory at the Tachyonic
Vacuum}, J.Math.Phys. {\bf 42} (2001)2844-2853, hep-th/0010240.


\bibitem{Branium}C.P. Burgess, P. Martineau, F. Quevedo and R.
Rabadan, \textit{Branonium}, JHEP {\bf 0306} (2003)037,
hep-th/0303170.
\bibitem{Felder}G. N. Felder, L. Kofman and A. Starobinsky,
\textit{Caustics in tachyon matter and other Born-Infeld scalars},
JHEP {\bf 0209}(2002)026, hep-th/0208019.\\
G. N. Felderand L. Kofman, \textit{Inhomogeneous fragmentation of
the rolling tachyon}, Phys. Rev. D{\bf 70}(2004)046004,
hep-th/0403073.



\end{thebibliography}

\end{document}